\documentclass[%
 aip,
 amsmath,amssymb,
 reprint,%
]{revtex4-1}

\usepackage{graphicx}

\usepackage{hyperref}

\usepackage{dcolumn}
\usepackage{bm}

\usepackage[utf8]{inputenc}
\usepackage[T1]{fontenc}
\usepackage{mathptmx}
\usepackage{etoolbox}

\usepackage{siunitx}

\usepackage{xcolor, soul}

\usepackage{natbib}

\setcitestyle{numbers}
\setcitestyle{square}

\makeatletter
\def\@email#1#2{%
 \endgroup
 \patchcmd{\titleblock@produce}
  {\frontmatter@RRAPformat}
  {\frontmatter@RRAPformat{\produce@RRAP{*#1\href{mailto:#2}{#2}}}\frontmatter@RRAPformat}
  {}{}
}%
\makeatother
\begin{document}

\preprint{AIP/123-QED}

\title[]{Needle in a haystack: efficiently finding atomically defined quantum dots for electrostatic force microscopy}
\author{José Bustamante}
 \email{joseb@physics.mcgill.ca}
\affiliation{ 
Department of Physics, McGill University %
}%

\affiliation{ 
Departamento de Física, Universidad San Francisco de Quito 
}%

\

\author{Yoichi Miyahara}%
\affiliation{ 
Department of Physics, McGill University %
}%
\affiliation{ Department of Physics,
Texas State University
}
\affiliation{Materials Science, Engineering and Commercialization Program (MSEC), Texas State University}

\author{Logan Fairgrieve-Park}
\affiliation{ 
Department of Physics, McGill University %
}%

\author{Kieran Spruce}
\affiliation{ 
Electrical Engineering Department, University College London 
}%

\author{Patrick See}%
\affiliation{ 
National Physical Laboratory, United Kingdom
}%

\author{Neil Curson}%
\affiliation{ 
Electrical Engineering Department, University College London 
}%

\author{Taylor Stock}%

\affiliation{ 
Electrical Engineering Department, University College London
}%

\author{Peter Grutter}
\affiliation{ 
Department of Physics, McGill University %
}%

\date{\today}

\begin{abstract}

The ongoing development of single electron, nano and atomic scale semiconductor devices would benefit greatly from a characterization tool capable of detecting single electron charging events with high spatial resolution, at low temperature. In this work, we introduce a novel Atomic Force Microscope (AFM) instrument capable of measuring critical device dimensions, surface roughness, electrical surface potential, and ultimately the energy levels of quantum dots and single electron transistors in ultra miniaturized semiconductor devices. Characterization of nanofabricated devices with this type of instrument presents a challenge: finding the device. We therefore also present a process to efficiently find a nanometre size quantum dot buried in a $10 \times 10~\unit{\mm^2}$ silicon sample using a combination of optical positioning, capacitive sensors and AFM topography in vacuum.

\end{abstract}

\maketitle

\begin{quotation}
\end{quotation}
\section{\label{sec:intro} Introduction}

Scanning Probe Microscopy (SPM) such as Scanning Tunneling Microscopy (STM) \cite{10.1063/1.92999} and Atomic Force Microscopy (AFM) \cite{PhysRevLett.56.930} are techniques that have contributed greatly to the development of nanoscience \cite{RevModPhys.75.949}. These techniques rely on utilizing the tunnelling current or the force between the tip and the sample to maintain precise control of the tip separation and measure local properties of the sample. Such techniques enable experiments on a single nanoscopic entity, such as a quantum dot, molecule or even an atom. With miniaturization, a non trivial problem arises: to find the sample and position it precisely below the SPM instrument's tip to perform imaging or spectroscopy. A widespread strategy is to prepare millions of identical copies of an object (molecules or nanoparticles) and to spread them over a surface in such a way that one assures at least one of them would be in the scanning range of the microscope.  
\\

This approach is not always possible with prototype and experimental nanoelectronic devices. For instance, the technique of STM hydrogen-desorption lithography  is capable of atomic-scale patterning of arsenic \cite{doi:10.1021/acsnano.9b08943} or phosphorus \cite{PhysRevLett.91.136104} on a silicon surface. A patterned island of dopants is then a QD, which can be as small as a single dopant atom. The same technique can be used to build highly conductive layers that serve as gates to the QD. AFM offers the possibility of characterizing such devices even after they have been encapsulated with a thin epitaxial layer of silicon. The forces measured by AFM on these devices are influenced by the material several nanometers below the surface.  However, to study such an engineered device the AFM needs to precisely position the tip on top of it: it is the problem of finding a needle in a haystack.

There have been two solutions to this problem implemented in  AFMs. A. MacDonald et al. integrated an optical microscope made with a bundle of optical fibers to obtain an optical image of the cantilever position at any temperature \cite{10.1063/1.4905682}. Ermakov et al. reported an AFM that uses a Scanning Electron Microscope (SEM) to acquire a large scale topographic image and use it as a guide to position the AFM tip \cite{10.1063/1.1144627}. These solutions are not suitable for our system for a variety of reasons including the incompatibility of low temperature with SEM, cost and space restrictions.  We need to perform low temperature ($4.5~\unit{K}$) electrostatic Frequency Modulation (FM)-AFM measurements with single electron charge sensitivity, and sub-nanometer resolution, to interrogate the charge states of quantum electronic devices structures. There is no commercial instrument able to do this.

With the aim of studying STM hydrogen-desorption lithography devices, we have built a low temperature AFM that is capable of finding a single specific dopant or quantum dot on a macroscopic chip. The AFM operates in a super-insulated cryostat with liquid nitrogen or liquid helium to maintain low temperatures. Therefore, the search for the device needs to be fast and efficient because of the limited hold time of our dewar. 

 In this article we present our new instrument and the series of steps to find a target location efficiently. We demonstrate that we can rapidly find a specific target  by using lithographically defined markers and capacitive position sensing, and perform electrostatic force microscopy characterization with single electron charging sensitivity.

 \begin{figure*}[ht]
 \centering

\includegraphics[width=\textwidth]{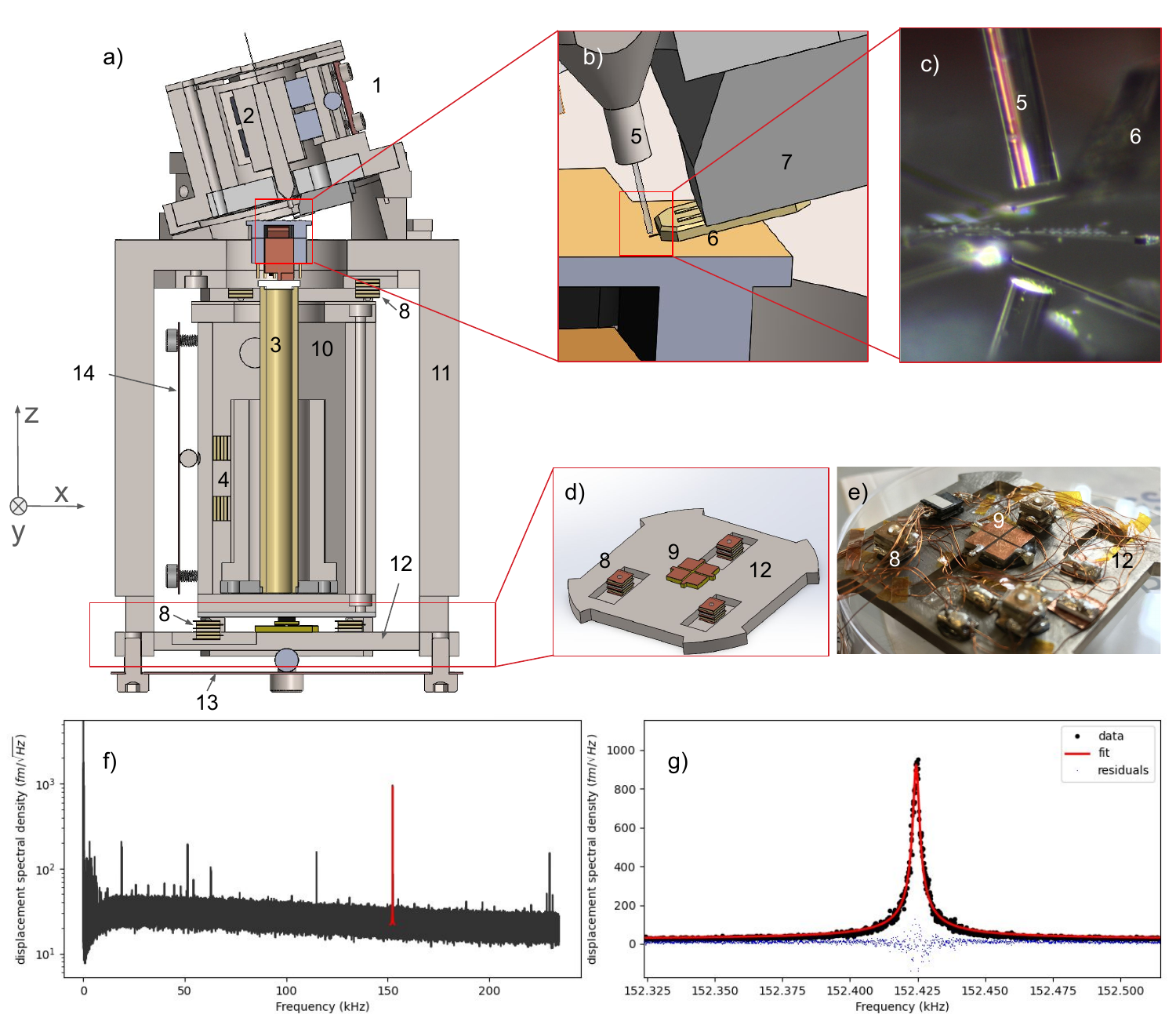}
\caption{\label{fig:cadfig} a) Physical design of the microscope. (1) Fiber walker. The fiber-cantilever cavity forms an Interferometer which is used to detect the displacement and excite the cantilever. (2) Movable hexagonal prism containing the ferrule and the fiber. (3) Scanner tube. (4) z-walker, 2 of the 6 piezo stacks are shown. (14) CuBe sheet for z-walker. All the z-walker enclosure (10) moves in xy. (13) CuBe sheet for xy-walker. (11) Microscope frame. b) Zoom to the fiber and the cantilever which need to be in close proximity. (5) Ferrule and optical fiber. (6) Cantilever chip is glued to a cantilever holder (7). c) Photograph showing the fiber close to the cantilever on top of the sample. The reflection can be seen on the sample. d)Zoom showing the xy-walker bottom plate.(8) xy piezo stacks. (9) Fixed electrodes of the position sensor. e) Photograph of the xy-walker bottom plate modelled in d). f) displacement spectral density of the microscope at 4.5 K with a g) zoom around the resonant frequency showing the thermal oscillation peak. A typical noise floor is  $\sim 18~\text{fm}/\sqrt{\text{Hz}}$.. }
\end{figure*}

\section{Instrumentation}

The heart of an AFM is the microfabricated silicon cantilever. We used a metal coated silicon cantilever (Nanosensors,PPP-NCLPt). The platinum iridium coating on both sides makes it conductive and prevents strong deformations of the cantilever as it cools down due to the bimetal effect. 

The microscope is designed to operate in the FM-mode \cite{10.1063/1.347347} at $4.5~\unit{K}$ (liquid helium), $77~\unit{K}$ (liquid nitrogen) and at room temperature by oscillating the cantilever and detecting both the frequency shift and oscillation amplitude (i.e. force gradient and dissipation). The  main challenge of a cryogenic instrument are the thermal expansion difference of different materials. First, two materials bonded with glue that experience a significant change in temperature contract or expand differently, and may delaminate. Second, the difference in thermal expansion can change the relative position of different parts of the instrument, compromising critical alignment. 

To mitigate these two effects (1) the choice of materials is driven by selecting materials with similar coefficient of expansion (2), a cylindrical symmetry is chosen to reduce changes in the position of the sample, which is located at the center of the cylinder,  and (3) the contraction of the microscope and resulting misalignment can be accommodated by  suitable sample coarse positioners.

In this design we choose to detect the cantilever displacement using an interferometer similar to the one used in reference \cite{Miyahara2020}. The key part of this set-up is that two different lasers are coupled into one fiber: one for opto-mechanically driving cantilever oscillations, the other for interferrometricaly detecting its displacement.  A displacement spectral density of the microscope when the cantilever is far away from the sample is shown in figure \ref{fig:cadfig} f) and g).  The typical noise floor of our AFM is measured to be $18 ~ \text{fm}/ \sqrt(\text{Hz}) $ at 4.5 K. This is enough to be able to distinguish single electron charging events tens of nanometers away from the surface.
  
For mechanical stability the cantilever is fixed with respect to the microscope body. The sample and optical fiber can be positioned using piezoeletric walkers. The cantilever chip is glued with a conductive silver epoxy (H20E, EPO-TEK \textregistered) to the cantilever holder which is screwed into the fiber walker. The whole fiber walker is at an angle of $\ang{15}$ with respect to the microscope body. The cantilever mechanical Q factors are typically 19k, 28k and 58k at $295~\unit{K}$, $77~\unit{K}$ and $4.5~\unit{K}$ respectively. Glueing the cantilever chip with the conductive epoxy results in a higher and more consistent Q factor compared to mechanically clamping the chip. When mounting the fiber walker, we align the end of the cantilever with the fiber by hand.
  
The sample is scanned by a piezoelectric tube scanner. To ensure that a reasonable sized sample area can be imaged and searched rapidly, the scanning range needs to be maximized at $4.5~\unit{K}$, hence also the piezo scanner tube length. Longer piezo tubes have a lower resonance frequency, leading to a greater susceptibility to mechanical noise. The microscope needs to fit in a 2'' bore of a superconducting magnet present in our cryostat. This magnet provides the possibility of performing experiments under 8 T magnetic field. We thus chose a 2'' long EBL2 piezoelectric tube (EBL products). Because of the temperature dependence of the piezoelectric constant this leads to the following xy scan ranges with 250V applied, at 4.5 K, 77 K and 295 K: $ \sim \SI{4}{\micro\metre} $  ,$10.4 \pm 0.5 ~\unit{\micro\metre} $, $11.4 \pm 0.5 ~\unit{\micro\metre}$. In the following we will describe some key aspects of our piezoelectric walkers and then report the implementation and characterization of the capacitive position encoders, both key components to finding a specific sample location rapidly.

\subsection{ Sample positioners }

The rest of the microscope is also designed with cylindrical symmetry.  We engineered a system of two stick slip motors to control the displacement of the sample in the xy plane as well as z. The z-sample walker is a stick slip motor to move the sample in the z direction, approach the sample and correct for any thermal changes to the tip-sample separation (recall the tip is fixed in space, only the sample and fiber can be stepped). The sample can also be moved laterally by the x-y sample walker. Typical stepsizes of these motors are ~50nm-100nm. Key to finding a given target rapidly are the capacitive position sensors that measure the z, x, and y coordinates of the sample walkers. We will now describe the sample walkers and the capacitive position sensors.

\subsubsection{z-sample walker}

The z-sample walker is a piezo-electric stick slip motor that displaces the sample, the scanner tube and its hexagonal housing (fig \ref{fig:cadfig} a) 4). It approaches or retracts the sample towards the cantilever in the z direction, the direction of motion is determined by the asymmetry of the sawtooth voltage applied.  
It is composed of six home made piezo stacks that clamp the hexagonal scanner housing. Each piezo stack is composed of four layers of EBL2 shear piezo (EBL2 shear piezo plate 1.0” $\times$ 1.0 $\times$ 0.019”, EBL products). The plates were cut into 4mm $\times$ 4mm squares.  The top surface of the piezo stacks are alumina plates which are in contact to sapphire plates(Swiss Jewel Company). The sapphire plates are glued to the hexagon housing and slide over the alumina surfaces. 

The choice of glues is critical for the reliability of the microscope. All layers of piezo stacks were glued with the conductive silver epoxy H20E . The piezo stacks themselves and the sapphire plates on which they walk, were glued to Titanium using Stycast 2850MT (Stycast 2850MT, Loctite \textregistered ), an electrically insulating general purpose encapsulant. 

The walker needs to be spring loaded to provide the normal force necessary for stick-slip motion. This is done through one of the sides of the hexagon. As shown in fig. \ref{fig:cadfig} a) four piezo stacks are glued to the outer z-walker enclosure (10), and two are glued to a z-walker plate. This plate is spring-loaded using a 3mm diameter sapphire sphere and a CuBe rectangular sheet. This configuration mitigates thermal contraction effects and ensures reliable operation at the whole operating range of temperatures. A similar piezo walker controls the cantilever-fiber interferometer spacing.

\subsubsection{xy-sample walker} 
The xy-sample walker allows this AFM to displace the sample laterally on the xy plane. It enables the xy navigation capability of the microscope. 

Consider fig.\ref{fig:cadfig} a). The xy-walker displaces the z-walker enclosure (10) laterally. The xy-walker has six piezo stacks (8), three on top and three on the bottom. They clamp  the z-walker enclosure.  The piezo stacks for xy displacement have a sapphire semi-sphere on top, which contacts a large alumina surface glued to the z-walker enclosure. The box movement is constrained by the contact of the upper surface of the z-walker enclosure to the 3 top piezo stacks. This constrains 3 degrees of freedom, and the z-walker enclosure is free to move in x,y and $\theta$. The bottom plate (fig.\ref{fig:cadfig} a) 12) with 3 additional piezo stacks is free to addapt to the z-walker enclosure orientation. It is spring loaded on the bottom with a 0.2 mm thick CuBe sheet.

The piezo stacks used for xy displacement are composed of 4 shear piezo layers: 2 for x, and 2 for y. Even though our walker is not mechanically constrained to maintain a fixed $\theta$ (the angle from the x-axis to the y-axis), we have found that the walker does not rotate significantly. Constraining the walker in $\theta$ considerably increases the complexity of the motor \cite{Liu2022}.

\subsubsection{Position Sensors}

To encode the relative position of the xyz walker described in the pervious section, we used the capacitive sensors as  reported in ref. \cite{10.1063/1.1150656}.  The sensor is composed of 4 square electrodes fixed to the xy-walker bottom plate. An single movable electrode is placed on top and fixed to the z-walker enclosure. A small separation $d$ between the top electrode and the bottom fixed electrodes is crucial for the sensor performance, since the current signal is inversely proportional to $d$ (eq. \ref{eqn:current_sensors}). Effectively 4 capacitors are formed, and the value of these capacitances is directly proportional to the position relative to the top electrode. To measure the position, an AC voltage $V_j=V_0\text{sin}(\omega_s t+\phi_j)$ is applied to the $jth$ bottom electrode, with a phase difference $\phi_j=\phi_0+j \pi/2 $. A current $I_j=i\omega C V_j$ is induced into the movable top electrode. The in-phase and quadrature components of the current are proportional to the x and y positions of the sample following this equation:

\begin{figure}[h!]
\includegraphics[width=9.5cm]{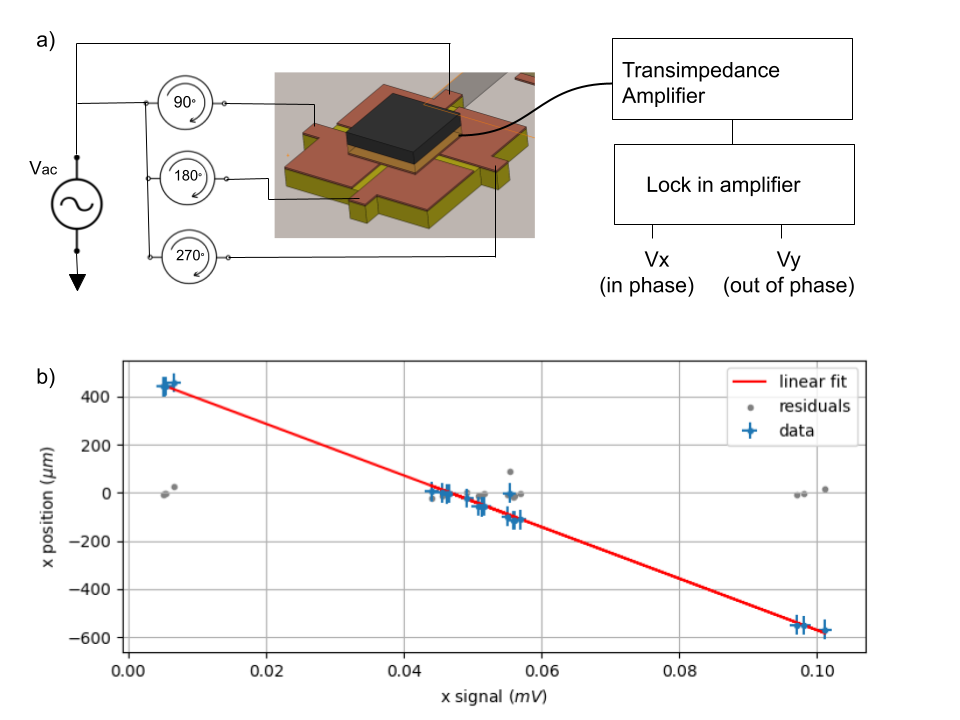}
\caption{\label{fig:sensor_cal} Capacitive xy position sensor. a) Schematic of the circuit and the CAD model of the 4 bottom electrodes, and the top movable electrode. A 10 Vpp, 10 kHz AC voltage is phase shifted and applied to each of the 4 fixed electrodes. The current transmitted to the top electrode is amplified in a transimpedance amplifier. The output voltage is amplified with a lock-in amplifier to obtain the in phase $V_x$ and out of phase $V_y$ components. b) Calibration of the sensors the position $x$ is a linear function of the voltage $V_x$. Uncertainty bars have been magnified: the plotted uncertainties in position are 10 times larger, and in voltage 100 times larger. }
\end{figure}

\begin{equation}\label{eqn:current_sensors}
I=2 \sqrt{2}l (\omega_s \epsilon_0 V_0/d)(x+iy)
\end{equation}

Where $I=Re \{ I \}+i Im\{I\}$ is the in-phase and out- of-phase components of the current, $l$ is the length of the electrodes, $\epsilon_0$ is the permittivity of free space, and \textit{x} and \textit{y} give the spatial position of the sample. Note that an overall phase $\phi_0$ results in a rotation of the x-y coordinates. It can be used to roughly align the sensor coordinate axis with the xy walker coordinate axis. 

To implement these sensors we used standard Printed Circuit Boards (PCBs) to fabricate the top and bottom electrodes. The bottom electrodes were soldered to kapton coated copper wire (California Fine Wire Company, MO232570)  while the top electrode was glued to a coaxial stainless steel cable (Lakeshore, Ultra miniature coaxial cable CC-SS-25), to reduce noise and stray capacitance. The bottom PCB bonded well to titanium with Stycast . However the top electrode detached regularly from the alumina surface when bonded with Stycast.   It bonded reliably to alumina with H20E.

The parameters chosen were the following: $2l=4~\unit{\milli\metre}$, $V_0=\SI{10}{\volt}$, $\omega_s=\SI{10}{\kilo \hertz}$, $ \phi_0=\ang{22}$. Given that $I \propto 1/d$ we chose a separation of $d=100 ~\unit{\micro\metre}$. Using these parameters we expect a capacitance of $ \SI{0.35}{\femto\farad}$ between the top electrode and a single one of the bottom electrodes when the top electrode is centered, with an expected current of $\SI{0.22}{\micro \ampere}$. 

  A separation of $100 ~\unit{\micro\metre}$ was achieved by covering the bottom electrode with an adequate layer of double sided Kapton tape, positioning the top electrode on top of the tape. This holds the electrodes in place during gluing. H20E can now be added to the upper surface of the top electrode and the z-walker enclosure is carefully positioned in its working position. The glue will absorb any height differences and ensure the desired separation of the plates. A hot air gun was used to cure the glue H20E for 30 minutes at 150 $ ^o C $, while monitoring the temperature with a thermocouple to avoid piezo depolarization. Once the glue has been cured, a small amount of ethanol was sprayed over the kapton tape to facilitate its removal. The separation was measured after fabrication to be $d=96.5 \pm 0.5 ~\unit{\micro\metre}$.

The current was amplified by a current amplifier SR330 with a gain setting of $3 ~\unit{\micro\ampere/\volt}$ and then measured using a lock in amplifier SR830 at a bandwidth of 55Hz. The position sensor was calibrated using a sample grid with markers at known positions. We fit the data to a polynomial function to convert from voltage [mV]  to position [$ ~\unit{\micro\metre} $]. The results are

\begin{equation}
x_p=-10720.4 V_x  -9.2 V_y + 500.1
\end{equation}

\begin{equation}
y_p=-2545.5 V_x -10273.6 V_y  -1273.8
\end{equation}

Where $x_p,y_p$ are the position x and y in $ \unit{\micro\metre}$, $V_x,V_y$ are the in and out of phase voltages in $\unit{V}$. Notice that while the $y_p$ position depends strongly on $V_x$ and $V_y$, x only depends strongly on $V_x$. We attribute this asymmetry to the top electrode's cable, which passes near the bottom electrodes and has an undesired capacitive coupling. For future designs the connection to the top electrode should come directly from the top to avoid this problem.

We conclude that the implemented capacitive position sensor has a resolution and accuracy of $0.5~\unit{\micro\metre}$, sufficient for our purposes as it is well within the xy scan range even at $\SI{4.5}{\kelvin}$.

\subsection{Sample holder with Temperature control}
\label{subsection:sample_holder}
The sample holder has to fulfill several boundary conditions simultaneously. It needs to position the sample in a fixed location with respect to the scanner, be mechanically stable, provide electrical connectivity, be easy to exchange, measure and control the temperature reliably.

\begin{figure}[h]
\includegraphics[width=8.5cm]{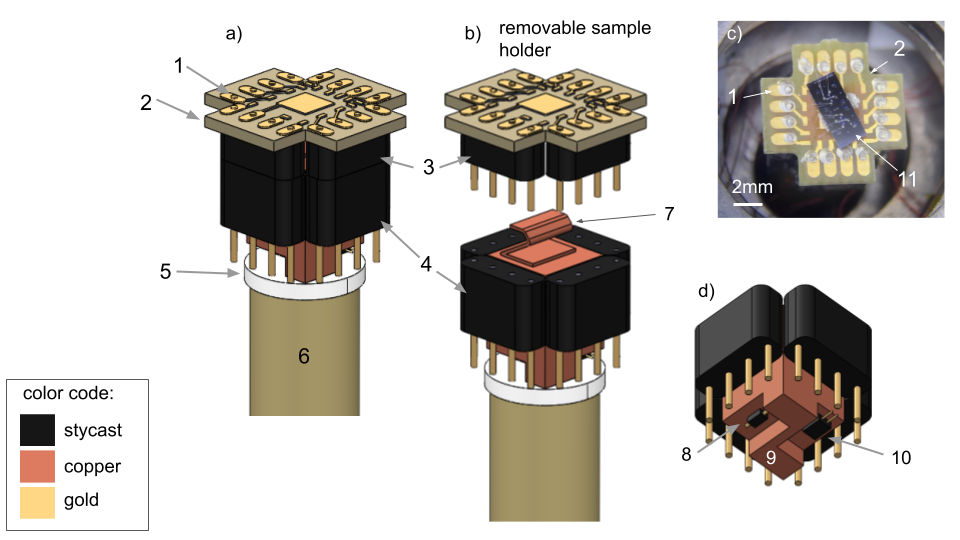}
\caption{\label{fig:sample_holder}  Sample holder with electrial connections and tempereture control. a) PCB (2) with metal contacts (1) to wirebond the sample. The PCB is soldered to four connectors (3) that are inserted into fixed receptacles (4). These receptacles are glued to a  a heat buffer (9) which  thermally stabilizes the sample. A macor spacer (5) electrically insulates the sample from the scanner tube (6). b) The PCB and connectors constitute the removable sample holder. A copper spring (7) increases thermal conduction from the heat buffer (9) to the PCB (2). c) Photograph of the sample holder with a silicon sample (11) bonded on top. The sample is glued to the center pad and wire bonded to the metal contacts on the surface of the PCB. d) Bottom view of the heat buffer (9) showing the the heater (8)  and temperature sensor (10).   }
\end{figure}

  We fulfilled all these requirements by the sample holder depicted in fig. \ref{fig:sample_holder}. On the scanner tube, we sequentially place from bottom to top: (1) a macor plate to isolate electrically the scanner tube from the scanner electrodes which carry a high voltage ($ \pm 250~\unit{\volt} $). (2) A copper cube. This heat conducting element has two pockets to position a temperature sensor (Lakeshore, DT-670) and a heater resistor to measure and control the sample temperature, allowing temperature dependent studies. (3) On the sides of the copper cube we bond 4 receptacles (Mill-Max,851-93-050-10-001000) that act as mechanical and electrical connections to the removable sample holder. (4) 4 connectors (Mill-Max,	850-10-050-10-001000) are inserted into the receptacles which are soldered to a (5) custom made PCB. The PCB has gold coated pads to wire-bond wires to the metallic pads on the silicon chip sample. This sample holder can be shipped from lab to lab, in our case from the UK to Canada.

\section{Finding the nanofabricated device}

\subsection{Samples of interest}

\begin{figure}[hb!]
\includegraphics[width=8.5cm]{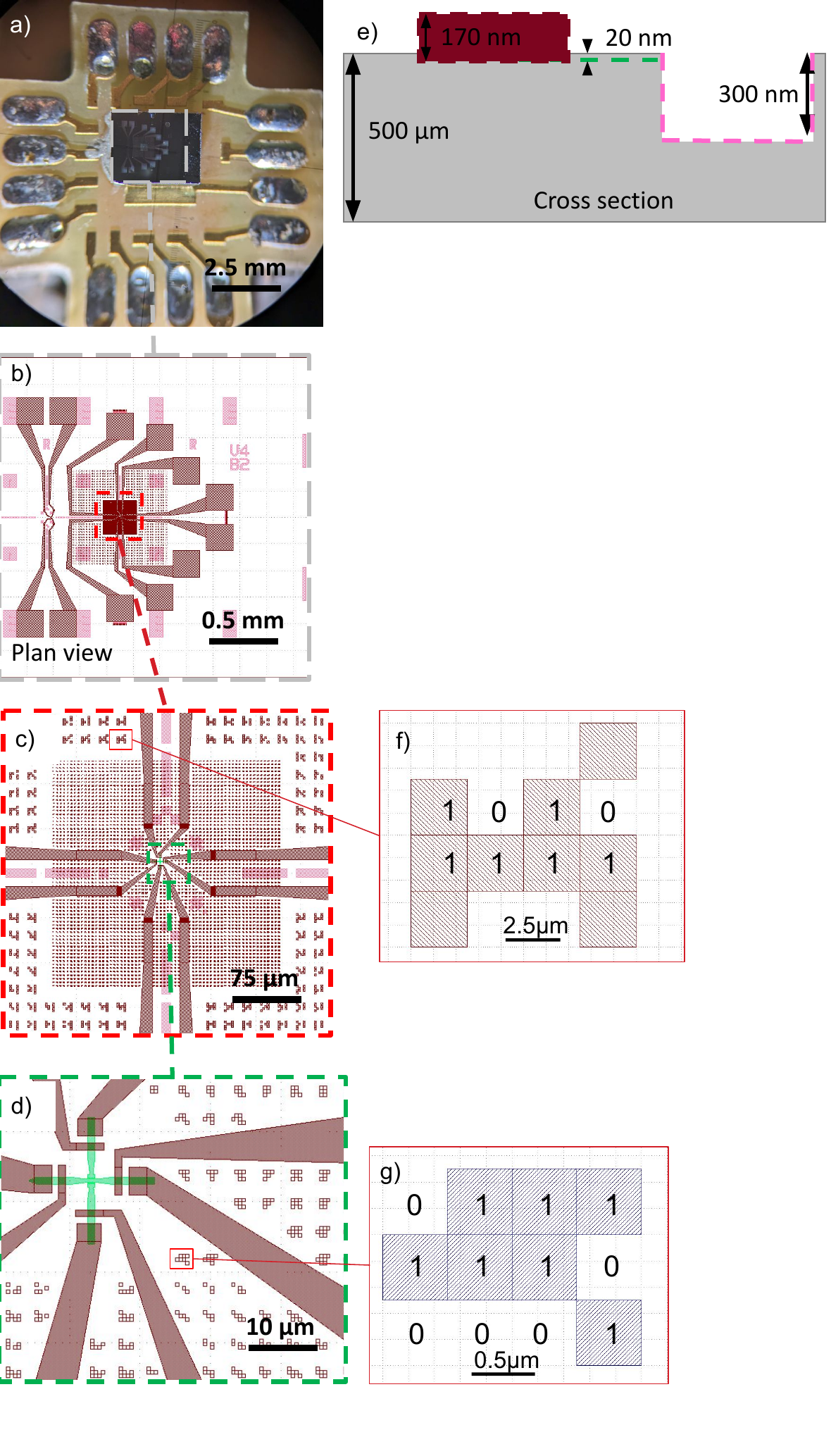}
\caption{\label{fig:digital_markers} Illustration of the sample from the sample holder to the QDs. a) Photograph of the sample mounted on the sample holder. b)GDS layout of the sample showing the bond pads and the large digital markers field. c) Some of the large digital markers  and the small ditigal markers field. d) target region showing the small digital markers, the metal electrodes connected with the dopant-defined electrodes in green. The QDs reside in the center of this cross. e) Cross section of the sample. Colors; red: metal features, pink: etched alignment markers for fabrication, green: dopant defined electrodes. f) $10\times 10~\unit{\micro\metre}$ outer markers for room temperature navigation. The absence of the top left square indicates that the marker is on that quadrant. g) Inner small markers for $4.5~\unit{K}$ navigation. 
  }
\end{figure}

The samples of interest are quantum dots and single electron transistors fabricated with the technique of STM hydrogen-desorption lithography as described in section \ref{sec:intro} \cite{doi:10.1021/acsnano.9b08943,PhysRevLett.91.136104}. In fig. \ref{fig:digital_markers} we illustrate the sample at its different length scales. The sample is fabricated on a $500~\unit{\um}$ thick Si chip shown in fig. \ref{fig:digital_markers} a), diced and bonded to the sample holder. The metal palladium electrodes are shown in red in the GDS layout with different magnifications (fig. \ref{fig:digital_markers} b-d). Digital position markers are also fabricated in the same step (fig. \ref{fig:digital_markers} f,g ).   Fabrication alignment markers are etched 300nm on the Si surface and are shown in pink. Fig. \ref{fig:digital_markers} e) is a cross section showing the heights of all these features.

\subsection{Definition of the problem and scanning strategies} \label{subsection:problem_definition}
We need to use this microscope to reliably find a $100~\unit{nm}$ area, on a macroscopic sample. An acceptable solution for this problem needs to be carefully defined. First, how fast do we need to find the target? The microscope operates at 3 different temperatures ($295~\unit{K}$, $77~\unit{K}$, $4.5~\unit{K}$) and the restrictions are more severe as the temperature decreases. At room temperature the time to find the target is solely limited by the patience of the operator. At $77~\unit{K}$ and $4.5~\unit{K}$ by the hold time of our dewar: 7 days and 3 days respectively. The majority of this time needs to be spent measuring the device and not finding it.  Therefore we impose a search time of less than 2 hours at $4.5~\unit{K}$, it can take up to a day at $77~\unit{K}$.  

There are other restrictions. Second, the risk of damaging the tip needs to be minimized during navigation, as we do not have the capability of in-situ tip exchange. The imaging strategy for navigation depends on the sample roughness. The samples we studied \cite{doi:10.1021/acsnano.9b08943} had very steep etched trenches 300nm deep and e-beam defined electrodes 170nm high.  The maximum scan speed of nc-AFM is slow when operating on such rough structures; it is easy to crash the tip.  Third, there is the unacceptable risk of damaging the sample, by either a mechanical crash or an electrostatic discharge. Our sample is unique and precious, the strategy to image it must ensure the sample survival. 

We have found the following solutions to these three problems.  To prevent tip damage we use the following scanning strategy. We scan far away from the surface to resolve large features using the electrostatic interactions. This was achieved by scanning in fm-AFM mode with tip-sample distance feedback control to keep a constant $ \Delta f$, with a large bias (5 to 9V) to the tip and a low $\Delta f \sim -5 ~\unit{\hertz}$ set-point. These parameters result in a large tip-sample z distance  ($\sim 400~\unit{\nano\metre}\text{ to }1 ~\unit{\micro\metre}$). Our z-scanner range is $1.6 ~\unit{\micro\metre} $, if we step the z-walker so that the scanner extension is at its midpoint, the tip-sample distance ($1 ~\unit{\micro\metre}$)  is larger than the maximum scanner extension ( $0.8 ~\unit{\micro\metre} $), hence a crash is very unlikely.   Scanning this far away results in poor lateral resolution, but we are interested in the large topographic features of the sample such as the metal fiducial markers and etched silicon trenches. Their heights are $+170 ~\unit{\nano\metre}$ and $-300 ~\unit{\nano\metre} $ respectively. These features are resolved well, rapidly and safely. (fig. \ref{fig:3strategies} b and c)

The risk of destroying the sample is partially addressed by avoiding tip crashes as described above. However, the sample can also be damaged by voltages applied to the electrodes. We will distinguish two kinds of threats: 1) electrostatic discharge (ESD) and,  2) currents that can induce electromigration \cite{PSHo1989}. Firstly, to avoid ESD while the sample is on the microscope we built a custom made connection box which will ensure connections between the sample and voltage sources are made with both sample and source grounded. To change a connection from one voltage source $V_1$ to voltage source $V_2$  we: 1) apply 0 V to the output of both voltage sources, 2) ground both voltage sources, 3) ground the contact on the sample, 4) disconnect the sample from source $V_1$ and connect it to source and the voltage source $V_2$, 5) disconnect the sample and the source from the ground. In this way, when we are physically changing connections, both the sources and the sample are grounded and protected against ESD. We have already discussed how to protect the sample while outside the microscope in section \ref{subsection:sample_holder}. Secondly, voltages on the ~1-5 V range applied to the delicate metal electrodes on the sample may induce electromigration of material and cause short circuits.  To prevent electromigration, we never apply more than a ~100mV voltage difference between any of the sample connections. While we are navigating the sample is grounded and electrically protected. 

In the following  we describe extensively how to find the target rapidly at low temperatures.

\subsection{ Procedure to find the target device}

We have developed  the following procedure to reliably and rapidly find the target of the device through a series of steps:

\begin{enumerate}
\item Wire-bond the sample on the removable sample holder. Ship the device.
\item Mount the device in the AFM sample holder. Optical preposition the cantilever on top of the device
\item Navigate towards the target. 
\item Find the QD on the STM defined buried electrodes. 
\end{enumerate}

The relevant details of this process are described in the following subsections. 

\subsubsection{Wire bonding and shipment} 

The devices we investigate by AFM are micro-fabricated in London, UK, by a combination of hydrogen-desorption lithography and conventional microfabrication techniques to generate macroscopic electrical contacts (as well as guidance markers, see below). To be characterized they have to travel 5200 km into our microscope. 

After the STM hydrogen-desorption lithography fabrication process \cite{doi:10.1021/acsnano.9b08943}, the devices are wire bonded to the custom removable sample holder (fig. \ref{fig:sample_holder} b). They are placed into a temporary receptacle that electrically connects all the terminals of the device to avoid ESD damage. They are shipped to Montréal in a container that protects them from ESD and mechanical shocks. 

\subsubsection{Optical prepositioning}

\begin{figure*}[ht!]
\includegraphics[width=\textwidth]{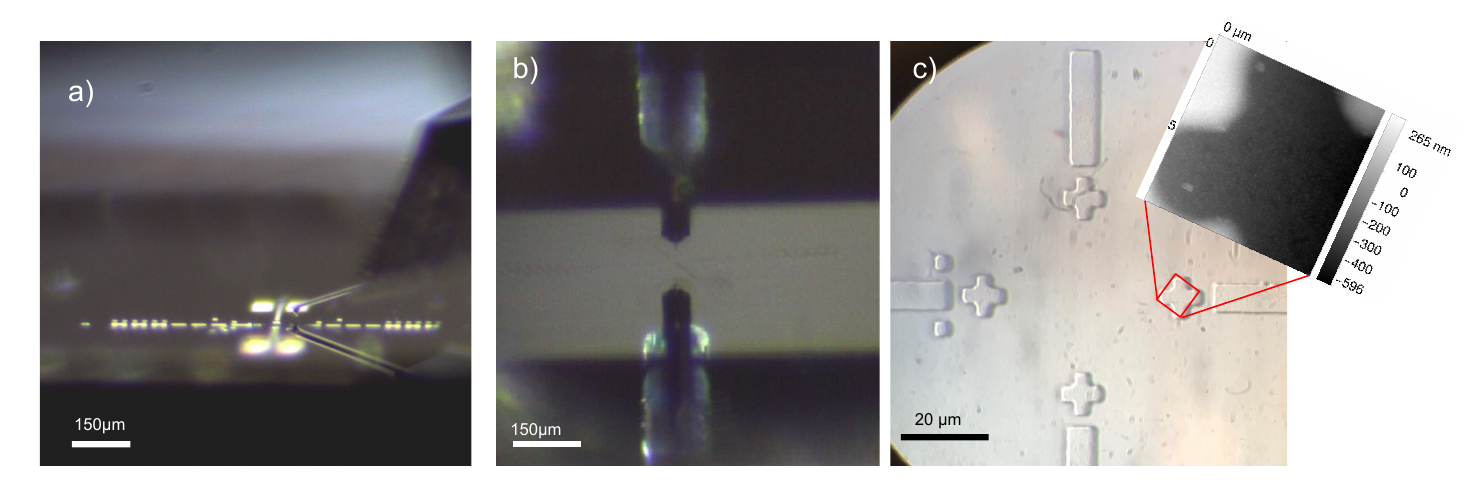}%
\caption{\label{fig:prepositioning} Positioning the sample with the aid of two perpendicular optical microscopes. a) Cantilever and tip seen from the side over a marker on the sample facilitates alignment on the x axis. b) Posterior view of the sample and the microscope to align on the y axis. c) Optical image of the sample markers with the AFM tapping mode scan taken immediately after optical positioning showing an alignment better than $\sim 10~\unit{\micro\metre}$.  
  }
\end{figure*}

In Montréal, the sample are carefully placed into the scanner sample holder during the process of assembling the AFM. The fiber walker with the cantilever is placed on top of the sample. It has three openings on the sides to allow optical access from the two sides of the cantilever and from the back. Using two long working distance optical stereo microscopes we can see the cantilever and the sample from two orthogonal directions simultaneously. This configuration avoids the difficulties arising from parallax when viewed only from one direction. While the accuracy one has on the axis perpendicular to the line of sight of the microscope is very good, estimating the depth is particularly hard. This problem is solved with the two perpendicular optical images. In fig. \ref{fig:prepositioning}  a) we can see the optical access showing the side of the cantilever and the tip while in b) the cantilever is seen from behind. We positioned the tip over one of the topographic crosses microfabricated on the sample as alignment markers, approached the sample to the tip and took an intermittent contact mode scan in air. The scan shows the cross with remarkable precision. This experiment shows that the target region of the sample can be optically positioned with a $\sim 10~\unit{\micro\metre}$ accuracy under the tip.

\subsubsection{Navigation towards the target}

After optically prepositioning the target location of the sample under the tip, we need to make sure the relative position is preserved. This is not trivial given the next steps necessary to cool down the microscope: sealing the vacuum can and transferring the microscope into the cryostat. A vacuum can is sealed with an indium wire that is uniformly squeezed using bolts. As we tighten the bolts, the whole microscope shakes. This motion can slightly move the sample with respect to the the tip. After the vacuum can is sealed, the microscope has to be inserted in the dewar to cool it down. We use a pulley and rail  system to carefully accomplish this, but again some mechanical bumping is unavoidable.

Even though we perform this process with great care, some misalignment is  inevitably introduced. Once in the can we need to re-position the sample to the original coordinates. In some cases the position changes by more than $\sim 100~\unit{\micro\metre}$. A major challenge is that a small rotation in the x-y sample walker, or changes in the stray capacitances of the position sensors introduce systematic errors in the position measurements. 

From the location where the tip landed, we need to navigate to the target device in the sample. We do this by taking successive scans in a chosen direction of the sample, until we find a marker. From the topography of the markers one can deduce the absolute location in a sample map. Once an absolute location is known, we can use the capacitive sensors to rapidly position the AFM tip over the target with the xy walker.

\subsubsection{Comparison of navigation strategies}

\begin{figure*}[ht!]
\includegraphics[width=\textwidth]{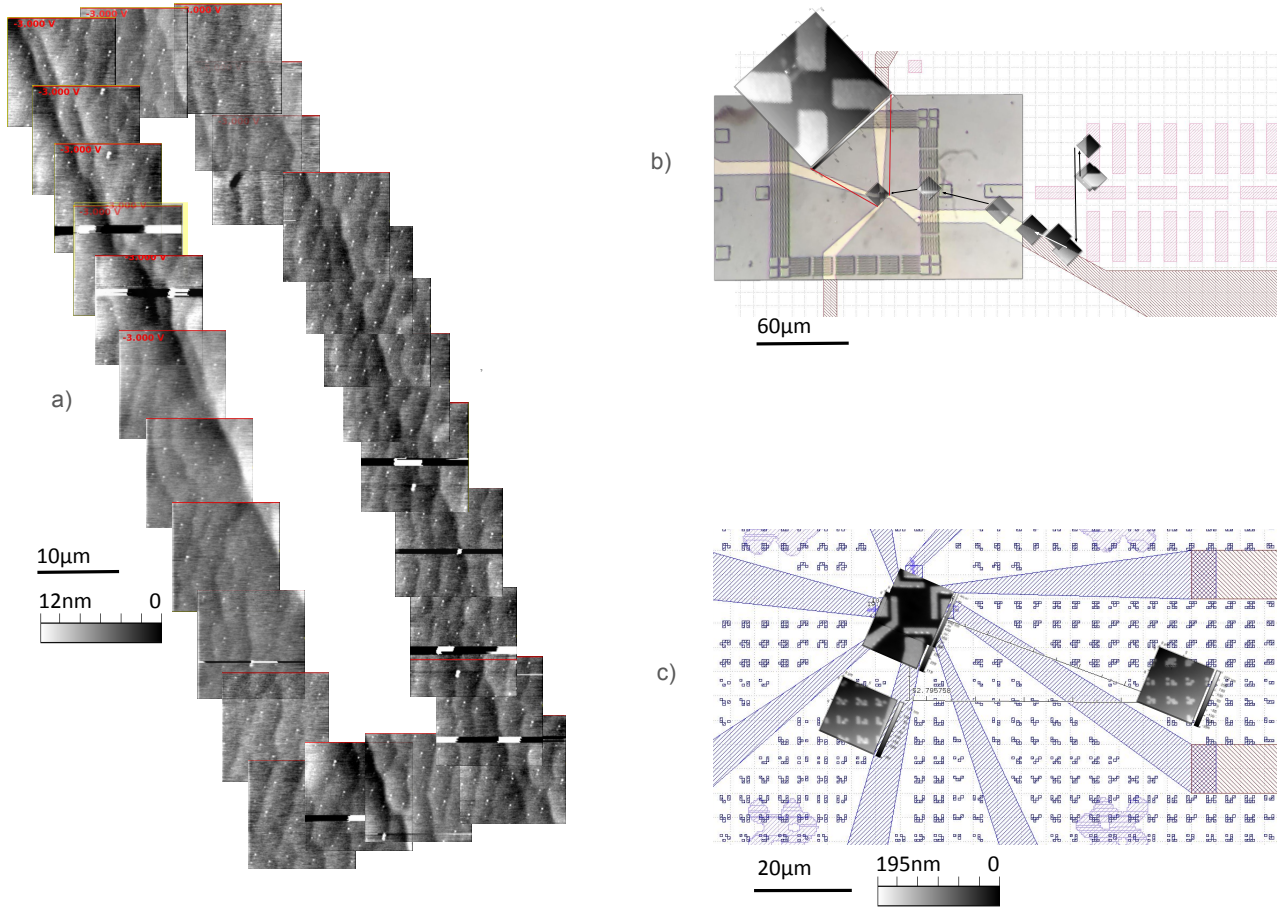}
\caption{\label{fig:3strategies} 3 different searching strategies. a) No sensor and no markers. Attempting to sweep large areas is not effective. Target not found. The tip is scanning $z  \sim 300~\unit{\nm}$ above the surface at $V_b=-3 ~\unit{\volt}$ and  $\Delta f= -5~\unit{\hertz}$ b) Position sensors and native markers. Using the position sensor described before and the alignment markers needed for micorfabrication of contact leads, allowed us to find the target. However, time consuming and not efficient. c) Using a position sensor, native markers and electrodes, and dedicated position encoding  markers optimizes finding the target device. In this device, the target was found with only 3 scans. Each AFM image uses all the scanning range of $11.4 ~\unit{\micro\metre} $ }
\end{figure*}

In the following we will describe the rational for the choice of digital markers. Fig. \ref{fig:3strategies} illustrates the different strategies we have tried to 'find the needle in the haystack' and how they compare to each other.  Initially, we tried to look for the target using a comprehensive scan of the sample without  position sensor or significant topographic markers. We prepositioned the tip to one side of the target and we systematically imaged a large area, as shown in \ref{fig:3strategies} (a). 

Interestingly, the Si step bunches are clearly visible even when scanning with the tip-sample distance of several hundred nanometres. The same step bunches are resolved in STM during fabrication. This suggest the possibility of using these inherent topographic features as a way of navigating and finding the target. While it is an interesting possibility, and a challenging problem of pattern recognition, we found that it is not efficient. These features are everywhere in the sample and it takes substantial effort to compare images taken with different microscopes and different conditions. The strategy of systematically scanning large areas did not work to find the target region. 

Subsequently, we installed the position sensor and we took advantage of the alignment markers used for sample fabrication to guide us towards the target. We were able to find the target. However, the markers were ambiguous so we needed many scans and a long time ($\sim$40 scans, $\sim$10 hours) to find the target. Hence, this strategy was not an acceptable solution to our problem as defined in section \ref{subsection:problem_definition}.

Finally, we designed digital markers that uniquely encode the position. This allows us to rapidly find the target with only 3 scans. The digital markers decreased the navigation time significantly.

\subsubsection{Digital position markers}

Even though it is possible to navigate using the etched microfabrication alignment markers and the metal electrodes of the sample as shown in fig. \ref{fig:3strategies} b), this procedure proved to be time consuming and ambiguous. Thus the need of unique fiducial markers that unambiguously encode the absolute position of the sample.

To facilitate navigation we designed digital markers that uniquely define the absolute position on the sample. The major design constraint was that within one scan area (which depends on temperature) we would be able to observe a marker and thus establish the absolute position on our sample. As shown in fig. \ref{fig:digital_markers}, we designed two sets of markers: large markers for room temperature, and small markers for cryogenic temperatures. A total of $710 \ \times \ 710 ~\unit{\micro\metre}^2$ is patterned with large markers, with a $256 \ \times \ 256 ~\unit{\micro\metre}^2 $ center patterned with small markers only. The markers were fabricated in the same step as the metal electrodes using ebeam lithography and lift-off. The height of these features was 170nm. The large marker field is designed with a large safety margin, we expect to never land outside of it.

At room temperature the scanning range is $11.4~\unit{\micro\metre}$. To ensure that a marker is imaged within two scans they must be $20~\unit{\um}$ apart, and have a density of  $20x20~ \unit{\micro\metre}^2$ per marker. If a total area of $(710 \times 710 -256 \times 256)   ~\unit{\micro\metre}^2$ needs to be uniquely encoded, $~ 1024=2^{10}$  different markers are required. This quantity can be encoded using 10 bits.

We implemented this encoding scheme by patterning  $10\times10~\unit{\micro\metre}^2$ markers on a $20~ \unit{\micro\metre}$ grid (fig. \ref{fig:digital_markers}).  Each marker consists of a grid of $4 \times4$ squares. Each square is $2.5~ \unit{\micro\metre}$ in size (fig. \ref{fig:digital_markers} b-f). The presence or absence of a square encodes a bit of information. Since we only require 10 bits, we encoded the marker number in the 8bits on the middle two rows, and we used the four corner squares to encode the four different quadrants of the sample. The absence of a square indicates the quadrant.

At $4.5~\unit{K}$ the scanner range is 4$~\unit{\micro\metre}$. To find one marker within this scan range with 100 \% reliability,  $4096=2^{12}$ different markers are needed for an area of $256 \times 256 ~\unit{\micro\metre} ^2$. Each marker thus consists of 3 rows of 4 squares encoding 12 bits. Each square measures $0.5 \times 0.5 ~\unit{\micro\metre}^2$ (fig. \ref{fig:digital_markers}g).  At $77~\unit{\kelvin}$ and at $4.5~\unit{\kelvin}$, the smaller marker field is sufficient to navigate efficiently.

\section{Finding a buried atomically thin electrical contact} 

Once the target region has been found, the described scanning strategy using long-ranged electrostatic interactions needs to change since the main feature of interest is not topographic anymore. We find there is no topographic signature on the surface that can be attributed to the  the dopant defined layers below. We present here several strategies to find the dopant-defined electrodes and thus find an atomic scale region of interest on a chip.

\subsection{Electrostatic force microscopy}

\begin{figure}[h]
\includegraphics[width=8.5cm]{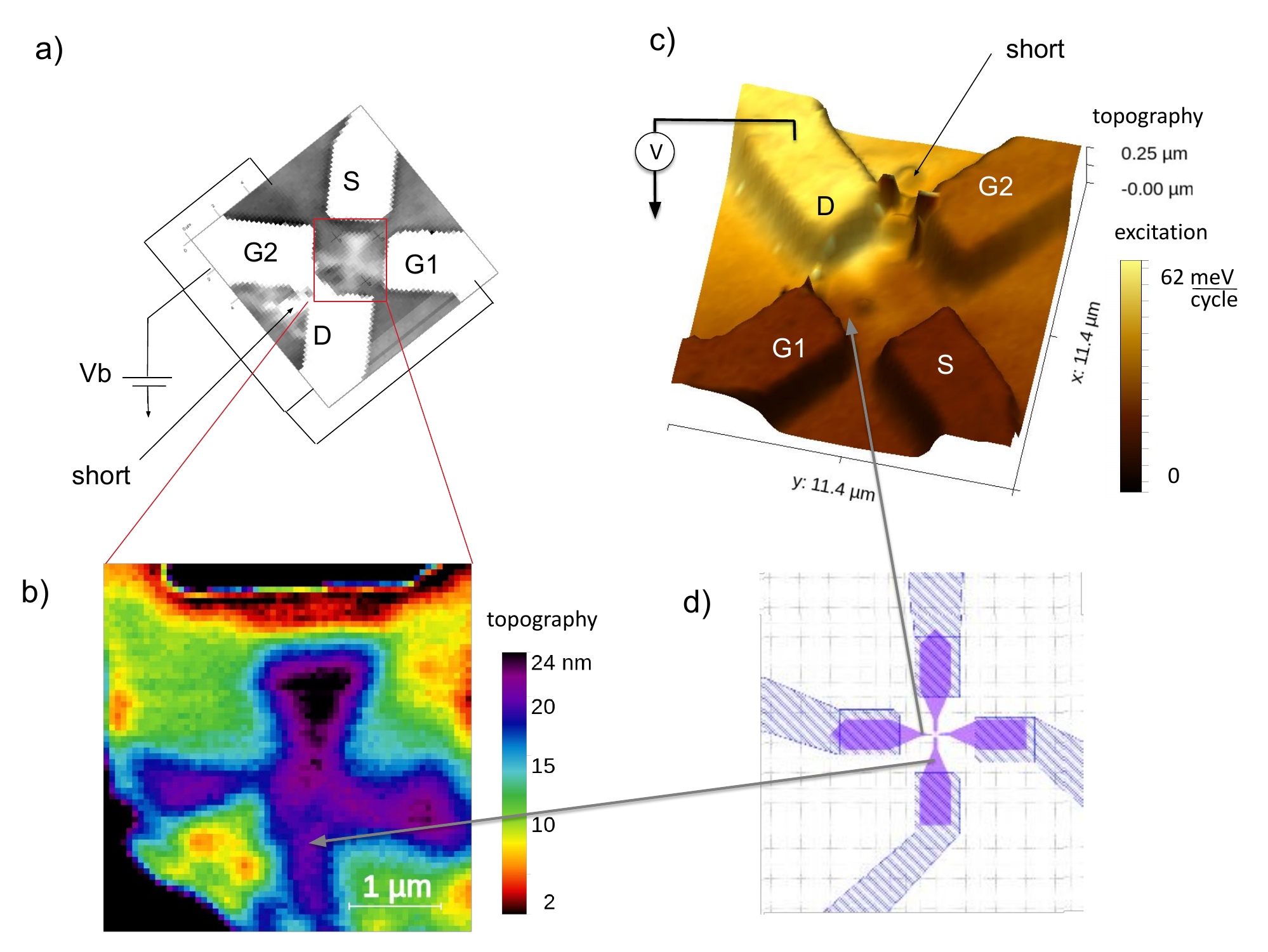}
\caption{\label{fig:dc_ac_efm} Electrostatic force microscopy used to characterize dopant defined STM electrodes. a) EFM image of the sample as 4V is applied to all the terminals. The device with its metallic electrodes and b) a zoomed image of the target showing the STM contacts. The height of the tip is shown as a function of position. c) Map of the electrostatic force produced by an AC voltage applied to one of the electrodes. Topography of the sample in 3D. The color scale is the excitation channel. The voltage applied is $V=V_b+\text{sin}(2 \pi f_0 t +\pi/2)$.  Notice how the excitation increases where there is an AC potential applied. There is evidence of a short circuit as there some of the potential applied to D is detected in  G2.   d) GDS layout of the design location of the metal electrodes and STM defined electrodes. 
}
\end{figure}

For the electrostatic forces measured on the present samples, non-contact, frequency modulated AFM, senses the force gradient as a change in the frequency shift of the oscillating cantilever. As a voltage is applied to the sample electrodes the changes of the electrostatic landscape can be measured with nc-AFM. The electrodes of the sample are metallic over the silicon surface. These metallic electrodes can be seen in white in fig. \ref{fig:dc_ac_efm} a) and in hatched blue in fig. \ref{fig:dc_ac_efm} d).They are connected to an atomically thin dopant layer that serves as a gates, source and drain,  fabricated with STM hydrogen desorption lithography. The layout of these STM contacts can be seen in fig. \ref{fig:dc_ac_efm} d) in purple. 

In the following, we demonstrate the detection and characterization of the  STM defined contacts buried $20~\unit{nm}$ below the surface of the sample. In fig. \ref{fig:dc_ac_efm} a) and b) , we observe contrast in the z channel as we scan with constant frequency shift. A voltage of 4V is applied to all the sample contacts (except S which is floating) while the tip is grounded. This image (fig. \ref{fig:dc_ac_efm} b) reveals the buried dopant defined contacts and clearly shows the central region where a quantum dot or single dopant would be located. We observe a dark region only on the contact that is floating.

We can use electrostatic force microscopy to characterize the connectivity of the electrodes. We observe a feature that resembles a short circuit between two of the metallic contacts G2 and D. This suggests that there might be an electrical connection between them.

Applying a phase locked  AC voltage $ \ang{90}$ out of phase with respect to the oscillation of the cantilever causes increase in the apparent damping of the cantilever which can  be detected directly in the excitation channel of the microscope (often called dissipation channel in the nc-AFM community). Note that this is a direct electrical drive, similar to the method used for measuring the transfer function of nc-AFM \cite{PhysRevB.84.125433} and dissipation-modulated Kelvin probe force microscopy \cite{PhysRevApplied.4.054011,miyahara2017force}. This AC voltage drives the cantilever at resonance using the electrostatic force. As this electrostatic driving force is at resonance the AFM is particularly sensitive to small interactions. 
We chose to apply a signal with a 6mV amplitude which is phase-locked and $\ang{90}$ phase-shifted from cantilever oscillation. In this way, the frequency shift can be used to track the topography and the effect of the AC applied electrostatic force appears in the excitation channel only. This allows us to correlate topography (e.g. of an electrode) and the presence of electrical signals applied to it simultaneously. Miyahara et al. have shown that the excitation signal reflects the surface potential of the sample\cite{PhysRevApplied.4.054011}. Here we are able to extract the contrast from the presence of this force on the electrodes or its absence on the rest of the sample.

In fig. \ref{fig:dc_ac_efm}c) we observe how the surface potential changes spatially.  We detect the AC potential on the metal electrodes as well as on the on the dopant defined electrodes. This AC method reveals the location of the quantum dot with higher precision than the DC polarization of the electrodes presented in fig. \ref{fig:dc_ac_efm} b). The correlation between the topographic bridge between D and G2, and the change in excitation of G2 caused by the signal applied to D can be clearly seen. We interpret this due to a high resistive electrical link between D and G2.

\subsection{ Single electron charging events indicate the location of quantum dots and traps}

A potential method to find QDs uses the very fact that they can be charged. Charging events are detected as ring structures on the frequency shift and / or dissipation channels \cite{cockins2010energy}. The QD is located at the center of these rings. This technique, called e-EFM, is described in the following section in which we detail the use of e-EFM to locate charge traps in the SiO$_x$/Si interface of our test sample.

When the surface of the silicon chip is in contact with oxygen, it is oxidized creating a silicon-silicon oxide interface. This Si-SiO$_x$ interface has a variety of defects such as interfacial traps $P_{b0}$ and $P_{b1}$ \cite{PhysRevB.73.073302,Megan2023}. The energy of these traps is inside the band gap of silicon. The tip, vacuum gap and silicon behaves as a metal-insulator-semiconductor (MIS) capacitor. A voltage applied to the tip is analogous to a gate voltage on a MIS capacitor which causes a charge reorganization (i.e band bending). For the n-type semiconductor we use, a positive gate voltage causing accumulation of charges and the conduction band bends downwards towards the Fermi energy. Eventually, an electron in the conduction band has a the same energy as the interfacial trap, hence tunnelling to the trap is favourable. The mechanical oscillation of the cantilever causes an oscillation of the conduction band location, so that an electron stochastically tunnels in and out of the trap, given the adequate voltage.    The in-phase component of this oscillating force is detected in the frequency shift channel and the out of phase component in the excitation channel. If the voltage is higher than this resonance, the trap is permanently charged \cite{Miyahara_2017}.

Scanning the tip laterally with a fixed applied voltage changes the band bending of a specific trap. The band bending is highest as the tip is exactly on top of the trap, and decreases as the lateral distance increases. A ring-like feature then appears at the locations where the tip induces an alignment of the trap energy level and the conduction band at the trap location. 

\begin{figure}[h!]
\includegraphics[width=8cm]{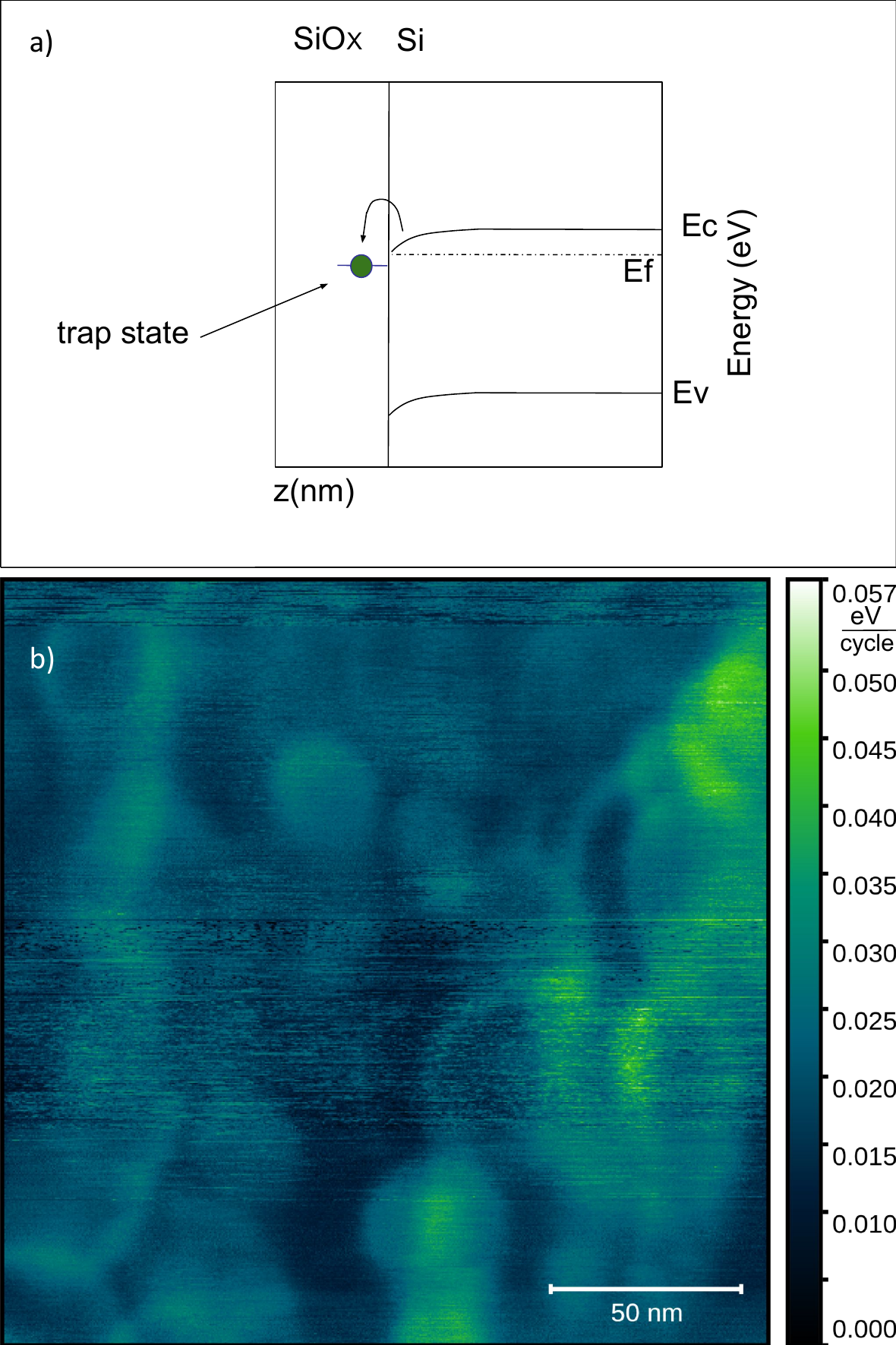}
\caption{\label{fig:rings_fairloop}  a) When the tip-sample potential changes, the semiconducting sample responds by band bending \cite{sze2012semiconductor} \cite{Megan2023}.The bands can bend sufficiently to charge interfacial traps if the tip-sample potential is above a threshold. This threshold potential is related to the trap energy level $qV_{trap}$. The charging of a trap is detected by e-EFM as a change in resonance frequency and dissipation \cite{Miyahara_2017}.(b) Excitation on the target region. Rings show the position of traps. The potential at a particular location on the sample can be changed in two different ways: either the tip-sample bias $V_b$ is changed at fixed tip-sample position or, alternatively, the tip is moved closer to the trap at a fixed tip-sample bias $V_b$. The image shows a scan at a fixed tip-sample bias of  $V_b = 8~\unit{V}$ at  $T =77~\unit{K}$. When scanning with a constant $V_b$, the potential at a trap is a function of the tip position x and y. A ring thus corresponds to the tip position where the effective potential at the trap location enables ionization of the trap. Different traps have different energies, leading to different ring diameters. The ring density enables the trap density to be measured.
 }
\end{figure}

In fig. \ref{fig:rings_fairloop} b), we observed charging rings at $77~\unit{K}$ over the the center of the target region. Applying the e-EFM technique to this sample reveals these acceptor like states that can be occupied by an electron. This data conclusively demonstrates that the AFM has enough sensitivity to perform single electron bias spectroscopy on quantum dots and provides a method to precisely locate these QDs.

\section{Conclusions}

We show an efficient method to find nanofabricated devices using a combination of optical alignment, capacitive encoded coarse positioning of the AFM and micorfabricated digital markers to reduce the search time considerably. This method will allow the researcher to concentrate on performing characterization experiments of the devices instead of finding them. We demonstrate how electrical excitation can be used to image buried dopant-defined electrodes and hence show the location of a particular device. Furthermore electrical excitation can be used to characterize the electrical continuity of of electrodes. Finally, we have introduced a novel AFM system with enough sensitivity to detect single electron charging of single traps at the Si-SiO$_x$ interface. The charging events reveal the position of such traps and hence it is a method to locate them. This instrument opens interesting possibilities for interrogating single electron, nano and atomic scale semiconductor devices.

\section{Acknowledgements}
 The authors would like to acknowledge funding from the following entities. Canada: National Research Council of Canada, Fonds de Recherche Quebec Nature Technologies, Institut Transdisciplinaire d'Information Quantique. United Kingdom: Engineering and Physical Sciences Research Council [grants EP/R034540/1, EP/V027700/1, and
EP/W000520/1] and Innovate UK [grant 75574]. Ecuador: Secretaría de Educación Superior, Ciencia, Tecnología e Innovación. The data that support the findings of this study are available from the corresponding author upon reasonable request.

Also, we would like to thank C. Boisvert, F. Belair, N. Sullivan-Molina, J. Smeros. R. Gagnon and R. Talbot for their support. 

The present article has been submitted to Review of Scientific Instruments. After it is published, it will be found at https://pubs.aip.org/aip/rsi.

\textcopyright~ Copyright (2024) José Bustamante. 

\section{References}

\nocite{*}
\bibliography{nltafm_rsi.bib}

\end{document}